\documentclass[lettersize, journal]{IEEEtran}
\usepackage{setspace}
\usepackage{cite}
\usepackage{pgfplots}

\pgfplotsset{compat=1.10}
\usepgfplotslibrary{fillbetween}
\usepackage{bm}
\usepackage{lipsum}
\usepackage{makeidx}
\usepackage{enumerate}
\usepackage{color}
\usepackage{cite}
\usepackage{amsmath,amsthm}   
\usepackage{amssymb}
\usepackage{nomencl}
\usepackage{multirow}
\usepackage{graphicx}
\usepackage{epstopdf}
\usepackage{threeparttable}
\usepackage{multicol}
\DeclareGraphicsExtensions{.pdf,.jpeg,.png,.jpg,.emf,.eps}
\usepackage{calc}
\usepackage{here}
\usepackage{url}

\usepackage{upref}
\usepackage{comment}
\usepackage{times}
\usepackage{dsfont}
\usepackage{epic,eepic}
\usepackage{rawfonts}
\usepackage[T1]{fontenc}
\usepackage{latexsym}
\usepackage{amsfonts}

\usepackage[font=small,labelfont=bf]{caption}
\usepackage[font=small,labelfont=bf]{subcaption}

\usepackage{xcolor}
\usepackage{tikz,pgfplots}
\usetikzlibrary{calc}
\usetikzlibrary{intersections}
\usetikzlibrary{arrows,shapes}
\usetikzlibrary{positioning}

\newtheorem{lemma}{Lemma}

\newtheorem{remark}{Remark}

\theoremstyle{definition}

\allowdisplaybreaks

\begin{document}
\title{RIS-Aided RSMA Improves the Latency vs. Energy Trade-off in the Finite Block Length MIMO Downlink}

\author{Mohammad Soleymani, \emph{Senior Member, IEEE},  
Bruno Clerckx, \emph{Fellow, IEEE}, 
Robert Schober, \emph{Fellow, IEEE}, and
Lajos Hanzo, \emph{Life Fellow, IEEE}
 \thanks{ 
Mohammad Soleymani is with the Signal and System Theory Group, University of Paderborn, 33098 Paderborn, Germany (email: \protect\url{mohammad.soleymani@uni-paderborn.de}). 

Bruno Clerckx is with the Department of Electrical and Electronic Engineering, Imperial College London, SW7 2AZ London, U.K. (e-mail: \protect\url{b.clerckx@imperial.ac.uk}).


Robert Schober is with the Institute for Digital Communications, Friedrich Alexander University of Erlangen-Nuremberg, Erlangen 91058, Germany (email: \protect\url{robert.schober@fau.de}).

Lajos Hanzo is with the Department of Electronics and Computer Science, University of Southampton, SO17 1BJ Southampton, United Kingdom (email: \protect\url{lh@ecs.soton.ac.uk}).

Robert Schober’s work was supported by the Federal Ministry for Research, Technology and Space (BMFTR) in Germany in the program of ``Souverän. Digital. Vernetzt.'' joint project xG-RIC (Project-ID 16KIS2432). The financial support of the following Engineering and Physical Sciences Research Council (EPSRC) projects is gratefully acknowledged: Platform for Driving Ultimate Connectivity (TITAN) (EP/X04047X/1; EP/Y037243/1); Robust and Reliable Quantum Computing (RoaRQ, EP/W032635/1); PerCom (EP/X012301/1); India-UK Intelligent Spectrum Innovation ICON UKRI-1859.
}}

\maketitle

\begin{abstract} We simultaneously minimize the latency and improve energy efficiency (EE) of the multi-user multiple-input multiple-output (MU-MIMO) rate splitting multiple access (RSMA) downlink, aided by a reconfigurable intelligent surface (RIS). Our results show that RSMA improves the EE and may reduce the delay to 13\% of that of spatial division multiple access (SDMA). Moreover, RIS and RSMA support each other synergistically, while an RIS operating without RSMA provides limited benefits in terms of latency and cannot effectively mitigate interference.  {Furthermore, increasing the RIS size amplifies the gains of RSMA more significantly than those of SDMA, without altering the fundamental EE-latency trade-offs.} Results also show that latency increases with more stringent reliability requirements, and RSMA yields more significant gains under such conditions, making it eminently suitable for energy-efficient ultra-reliable low-latency communication (URLLC) scenarios.
\end{abstract}
\begin{IEEEkeywords}
Energy efficiency, finite block length coding, latency versus energy efficiency trade-off, low latency, MIMO systems, rate splitting, reconfigurable intelligent surface.
\end{IEEEkeywords}

\section{Introduction}\label{sec:i}
Next-generation (NG) wireless networks have to support demanding applications, requiring low latency, hyper reliability, and high energy efficiency (EE) \cite{vaezi2022cellular}. In particular, the sixth generation (6G) of wireless systems is expected to improve EE and latency by factors of 100 and 10, respectively, compared to 5G networks \cite{wang2023road, gong2022holographic}. A powerful tool to achieve these ambitious goals is reconfigurable intelligent surfaces (RISs) \cite{wu2021intelligent, di2020smart}. Indeed, an RIS not only enhances the performance of ultra-reliable low-latency communication (URLLC) systems, but its benefits become more significant as the reliability and latency requirements tighten \cite{soleymani2024optimization, soleymani2024rate2}. Another promising technology to improve latency and EE is rate splitting multiple access (RSMA), which has been widely recognized for its effectiveness in managing interference \cite{mao2022rate}. In this paper, we investigate the latency-EE trade-off of RSMA-assisted RIS-aided downlink (DL) systems and quantify how the two technologies interact across different operational points and performance demands.

\subsection{Related Work}

A primary impairment degrading latency is interference. The optimal strategy for managing interference depends on its level of severity. When interference is weak, treating interference as noise (TIN) is the optimal approach. By contrast, when there is a single dominant interferer, detecting and subtracting it from the composite received signal is optimal. This approach is known as successive interference cancellation (SIC). Finally, in the presence of several similar power interferers, parallel interference cancellation (PIC) is the best solution. RSMA incorporates both TIN and SIC, making it more flexible and general than other multiple access schemes such as spatial division multiple access (SDMA), broadcasting, and multicasting \cite{mao2022rate, clerckx2023primer}. 

 In the RSMA DL, each base station (BS) transmits two types of messages: common and private. Each private message is intended for a single user, while common messages are detected by multiple users \cite{mao2022rate}. Several RSMA variants exist, differing in how common messages are structured and shared among users \cite{mao2018rate}. A practical and widely adopted RSMA architecture is the 1-layer rate-splitting (RS) scheme, which was first introduced in \cite{clerckx2016rate} and subsequently analyzed and optimized in \cite{joudeh2016sum}. The 1-layer RS relies on a single common message that is detected by all users in a cell, in addition to user-specific private messages. At the receiver side, each user first detects the common message, applies SIC to remove it, and then detects its intended private message while treating the remaining interference as noise. With a favorable balance between performance and complexity, the 1-layer RS architecture has emerged as a practical and widely adopted RSMA variant, and has been shown to outperform conventional SDMA and non-orthogonal multiple access (NOMA) across diverse deployment scenarios. In particular, significant gains in spectral efficiency (SE), EE, and robustness to imperfect channel state information have been reported for multi-antenna broadcast channels (BCs) and mobile environments in, e.g., \cite{mao2019rate, mao2018energy, dizdar2021rate, zhou2021rate}. Extensions to more general RS formulations, receiver architectures, and precoding strategies have further demonstrated the flexibility and practical relevance of RS-based transmission in beyond-5G and 6G systems \cite{li2020rate, flores2021tomlinson}.

Most of the aforementioned works, however, focus on single-stream transmission per user, which is well suited for single-antenna receivers or rank-one transmission scenarios. In general MU-MIMO systems where users are equipped with multiple antennas, restricting each user to a single data stream can be highly suboptimal \cite{soleymani2024optimization, soleymani2024rate}. Importantly, the benefits of RSMA are not limited to single-stream transmission and persist when multiple data streams are transmitted per user. In this context, RSMA has been extended to full MU-MIMO architectures with multi-stream transmission, where SE and EE gains over SDMA have also been observed \cite{mishra2021rate}. In this paper, we adopt the MIMO RS framework of \cite{mishra2021rate}, in which a vector of common streams is transmitted and detected by all users.

Another technology to enhance coverage and influence interference is RIS. Indeed, RISs can reshape the propagation environment, strengthening desired links and attenuating selected interference components \cite{huang2019reconfigurable, wu2019intelligent}. These capabilities have led to demonstrated performance gains in BCs and multi-user interference networks \cite{soleymani2023spectral, soleymani2025spectral, jiang2021achievable, santamaria2024interference}. However, RIS-based interference control is inherently constrained in DL cellular systems, where a single transmitter serves multiple users over a limited number of spatial data streams \cite{soleymani2022noma}. In such scenarios, RISs offer valuable additional degrees of freedom but cannot fully mitigate interference when the system load is high. Motivated by these limitations, RSMA has been combined with RIS to jointly enhance coverage and manage interference. The SE performance of RIS-aided RSMA was first studied in \cite{bansal2021rate}, showing clear gains over SDMA-based solutions in single-stream transmission settings with single-antenna base station and users. Further investigations revealed a synergistic interaction between RSMA and RIS, arising from their complementary roles in interference mitigation and channel enhancement \cite{li2023synergizing, li2022rate}, still under single-stream assumptions. These insights have also been extended to fully MIMO scenarios with multiple data streams per user, demonstrating that the RSMA-RIS synergy persists in multi-antenna systems and under practical hardware constraints \cite{soleymani2022rate, soleymani2023energy, soleymani2023rate}.

Finally, to meet stringent latency requirements, finite block length (FBL) coding is necessary, which limits the applicability of Shannon-based rate expressions \cite{polyanskiy2010channel}. The authors of \cite{dhok2022rate, pala2023spectral, katwe2024rsma, singh2023rsma, jorswieck2025urllc, wang2023flexible, xu2023max, xu2022rate} utilized RSMA to enhance the SE of multiple-input single-output (MISO) FBL BCs, both in scenarios with or without RISs. Moreover, the authors of \cite{katwe2022rate} revealed that RSMA increases the global EE of a MISO RIS-aided BC with FBL. As a further advance, the authors of \cite{soleymani2023optimization} illustrated that RSMA improves both the minimum rate and EE in a multi-cell RIS-aided MISO BC.  {More recently, RIS-assisted RSMA under FBL coding was also studied for a cell-free massive MIMO architecture with single-antenna users \cite{shen2025synergistic}, showing that RSMA enhances the weighted sum rate of the system. Note that the authors of \cite{dhok2022rate, pala2023spectral, katwe2024rsma,  wang2023flexible, xu2023max,singh2023rsma, xu2022rate, katwe2022rate, jorswieck2025urllc, soleymani2023optimization, shen2025synergistic} considered only single-stream transmission per user.} By contrast, RSMA supporting multiple streams was considered in \cite{soleymani2024rate} to improve the SE and EE in multi-cell MIMO URLLC RIS-aided BCs.

\subsection{Motivations and Contributions}
 The aforementioned treatises used RSMA for enhancing either the SE or EE of FBL systems. However, the salient latency-oriented performance metrics have remained largely unexplored, especially in RSMA-aided FBL systems. In particular, as far as we are aware, the minimum-maximum (min-max) delay of users has not been optimized for RSMA schemes, as highlighted in Table \ref{table-1}. Min-max delay is a critical metric for guaranteeing fairness and worst-case latency. Additionally, the joint latency-EE trade-off of RSMA has not been examined. Existing studies mostly consider SE or EE in isolation, overlooking the inherent interplay between ultra-low latency and energy-efficient communication, which is essential for NG URLLC.

These gaps are fundamental in the context of 6G performance requirements, where latency, reliability, and EE must be simultaneously improved \cite{wang2023road, gong2022holographic}. However, these objectives are naturally in contrast: reducing latency often demands higher transmit power and shorter codewords and packets, which reduce EE. Conversely, aggressively maximizing EE tends to increase delay due to a typically low transmission power. Therefore, understanding and optimizing the latency-EE trade-off is indispensable for providing realistic and implementable design guidelines, especially for RSMA and RIS architectures.

 \begin{table}
\centering
\scriptsize
\caption{Summary comparison of closely related studies on RSMA for RIS-assisted FBL systems.}\label{table-1}
\begin{tabular}{|c|c|c|c|c|c|c|c|c|c|c|c|c|c|c|}
	\hline 
 &This paper&\cite{soleymani2024rate}&\cite{soleymani2023optimization}&\cite{katwe2022rate}&\!\!\! {\cite{dhok2022rate, pala2023spectral, katwe2024rsma,singh2023rsma, shen2025synergistic}}
 \\
 \hline
 Ch. disp. in \cite{scarlett2016dispersion}&$\surd$&$\surd$& $\surd$& -&-
 \\
 \hline
 EE metrics&$\surd$&$\surd$&$\surd$&$\surd$&-
 \\
 \hline
 MIMO systems&$\surd$&$\surd$&-&-&-
 \\
\hline
Multiple streams&$\surd$&$\surd$&-&-&-
\\
\hline
Min-max delay&$\surd$&-&-&-&-
\\
\hline
EE-latency trade-off&$\surd$&-&-&-&-
\\
\hline

		\end{tabular}
\normalsize
\end{table} 

To address these limitations, this treatise closes the aforementioned knowledge gap by jointly investigating latency and EE in the RSMA-enabled MU-MIMO RIS-aided DL. Rather than focusing on a single operating point, we characterize the achievable EE-latency region and identify Pareto operating points that reflect diverse performance requirements. Additionally, we examine how RSMA and RIS interact across these operating points, shedding light on their roles and combined impact under different latency and energy demands. These insights form the basis of the contributions detailed in the following:
\begin{itemize}
    \item We formulate a new optimization problem for MU-MIMO RIS-aided systems that jointly minimizes the maximum user delay and maximizes the minimum EE, through a weighted objective capturing diverse EE-latency operating points. Although we leverage lower bounds for FBL rates from \cite{soleymani2024rate}, the resulting problem is entirely novel in the context of RSMA and RIS, and offers practically meaningful design insights that have not been explored before.

    \item  We demonstrate that RSMA and RIS fulfill complementary roles: \textit{RIS enhances link quality and coverage}, whereas \textit{RSMA is responsible for interference mitigation}. Consequently, their combination is synergistic, especially when the system operates in an interference-limited regime, namely when the number of users exceeds the number of BS transmit antennas. Indeed, in such scenarios, it is not feasible to serve all users over interference-free streams, and thus, an RIS alone cannot resolve the interference issue. The benefits of RSMA become more prominent as the number of users and the BS power budget increase, primarily due to the resulting higher interference levels. These findings reinforce the importance of RSMA as a flexible interference-management mechanism for NG multiple access.

    \item We analyze how reliability constraints, transmit power, and RIS size impact delay and EE. Moreover, we characterize the latency-EE performance of RIS and RSMA. In particular, our results show that:
    \begin{itemize}
        \item RSMA simultaneously decrease the maximum delay and enhance the minimum EE for all latency-EE operational points. 
        
        \item Improving EE typically increases delay, while reducing latency deteriorates EE, revealing a fundamental trade-off. This trade-off becomes more prominent as the BS power budget increases. 

        \item Latency increases with more stringent reliability requirements.  This is in line with plausible expectations, since a reduced bit error rate (BER) requires longer channel codes. 
        
        \item RSMA offers higher performance gains under tighter latency constraints, confirming its effectiveness for URLLC scenarios.
        
        \item Increasing the number of RIS elements enhances both latency and EE, but the magnitude of this improvement depends strongly on the multiple-access strategy and operating regime.
    \end{itemize}

\end{itemize}

\subsection{Paper Outline and Notations}
The paper is organized as follows. Section \ref{sec=ii} presents the system model, deriving the rate, EE, the delay expressions and the formulation of the optimization problem considered. Section \ref{sec:iii} designs an RSMA scheme for improving the latency-EE trade-off. Section \ref{sec-iv} presents numerical results, and Section \ref{sec-v} concludes the paper.  

 \textit{Notations}: ${\bf I}_N$ denotes the $N\times N$ identity matrix. Scalars, vectors, and matrices are represented by  lowercase (e.g. $t$), bold lowercase (e.g. ${\bf t}$), and bold uppercase letters (e.g. ${\bf T}$), respectively. Zero-mean complex proper Gaussian signal ${\bf t}$ is denoted by $\mathcal{CN}({\bf 0},{\bf T})$, where ${\bf T}=\mathbb{E}\{{\bf t}{\bf t}^H\}$ is the covariance matrix of ${\bf t}$. $\text{Tr}({\bf T})$, $|{\bf T}|$, and $\mathbb{E}\{{\bf T}\}$ denote the trace,  determinant, and  mathematical expectation of ${\bf T}$, respectively. Moreover, $\Re(t)$ is real value of $t$. Additionally, $Q^{-1}$ denotes the inverse of the Gaussian $Q$-function.

 \begin{figure}[t]
    \centering
           \includegraphics[width=.34\textwidth]{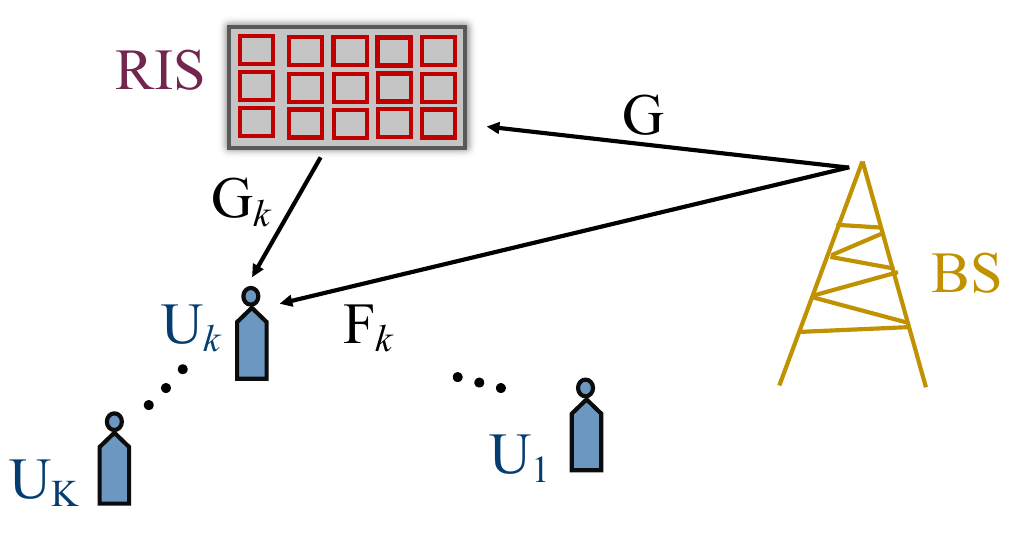}
    \caption{ {System model of the MU-MIMO RIS-aided downlink.}}
	\label{Fig-sys} 
\end{figure}
 
\section{System Model}\label{sec=ii}
We study an RIS-assisted MIMO BC, where a single BS equipped with $N_{BS}$ antennas communicates with $K$ multi-antenna users, each having $N_u$ receive antennas. The transmission is aided by an RIS comprising $M$ reflecting elements. The RIS configuration follows the modeling approach in \cite{soleymani2022improper,pan2020multicell}. Hence, the resulting channel linking the BS to user $k$ can be expressed as
$\mathbf{H}_{k}\left({\bf \Psi}\right)= 
{\mathbf{G}_{k}{\bf \Psi}\mathbf{G}}
+
{\mathbf{F}_{k}}
$,
where ${\bf F}_{k}$ denotes the direct channel from the BS to user ${k}$, ${\bf G}_{k}$ represents the channel matrix between the RIS and user $k$, ${\bf G}$ corresponds to the channel linking the BS and the RIS, as shown in Fig. \ref{Fig-sys}, while ${\bf \Psi}$  models the scattering response. A nearly passive RIS with a diagonal structure is assumed, i.e., ${\bf \Psi}=\text{diag}(\psi_1,\psi_1,\cdots,\psi_M)$, with $|\psi_m|\leq 1$ for all $m$.

 {
We assume the availability of perfect instantaneous global channel state information (CSI). This assumption is standard in RIS-assisted beamforming and URLLC resource allocation studies, as it enables the characterization of fundamental performance limits \cite{pala2022joint, ghanem2020resource, soleymani2024optimization}. In practical systems, RIS-related CSI can be acquired via pilot-based training, RIS reflection pattern design, and compressive sensing techniques \cite{zheng2022survey, wang2020channel, you2020channel}. The impact of imperfect CSI is an important direction for future resource-allocation studies.
}

\subsection{Signal Model and Rate Expressions}
We assume that the BS employs  the MIMO RS proposed in \cite{mishra2021rate}, implying that the BS transmits 
\begin{equation}
{\bf x}=
\underbrace{{\bf \Upsilon}{\bf s}_{c}}_{\text{Common M}}+
\underbrace{\sum_k{\bf \Upsilon}_{k}{\bf s}_{k}}_{\text{Private M}} \in \mathbb{C}^{ N_{BS}\times 1},
\end{equation}
where ${\bf s}_{c}\sim \mathcal{CN}({\bf 0},{\bf I}_N)$ is the common stream vector, ${\bf s}_{k}\sim \mathcal{CN}({\bf 0},{\bf I}_N)$ is the private stream vector intended for user ${k}$, while the matrices ${\bf \Upsilon}\in \mathbb{C}^{ N_{BS}\times N}$ and ${\bf \Upsilon}_{k}\in \mathbb{C}^{ N_{BS}\times N}$ are, respectively, the beamforming matrices for ${\bf s}_{c}$ and ${\bf s}_{k}$, where $N = \min(N_{BS}, N_u)$ is the maximum number of streams per user. The signals ${\bf s}_{c}$ and ${\bf s}_{k}$, consisting of maximum $N$ streams each, are assumed to be independent and identically distributed. We refer the reader to \cite[Section II.B]{mishra2021rate} and \cite[Section II.B]{soleymani2024rate} for further details.
We denote the set of beamforming matrices as $\{{\bf \Upsilon}\}=\{{\bf \Upsilon}_1,{\bf \Upsilon}_2,\cdots,{\bf \Upsilon}_K,{\bf \Upsilon}\}$.

In the MIMO RS proposed in \cite{mishra2021rate}, the messages of user $k$ are split into a common part, included in the vector stream ${\bf s}_c$, and a private part ${\bf s}_k$. Hence, user $k$ aims for detecting ${\bf s}_c$ and ${\bf s}_k$. The interfering signals ${\bf s}_j$, $\forall j\neq k$, are treated as noise by user $k$. Moreover, each user first detects ${\bf s}_c$, treating all other signals as noise. Then, the users apply SIC to the common part before detecting their corresponding private part. Therefore, the maximum achievable rate for ${\bf s}_{c}$ at user $k$ is \cite{soleymani2024optimization}
\begin{multline}
        r_{ck}\!=\!
\log\! \left|{\bf I} \!+\!\!\left(\!\!\sigma^2{\bf I}\!+\!\!\sum_j\!{\bf H}_{k}{\bf \Upsilon}_{j}{\bf \Upsilon}_{j}^H{\bf H}_{k}^H\!\!\right)^{-1}\!\!{\bf S}_{ck} \right|
\\
-Q^{-1}(\epsilon_c)\sqrt{\frac{2\text{Tr}({\bf S}_{ck}(\sigma^2{\bf I}\!+{\bf S}_{ck}+\!\!\sum_j\!{\bf H}_{k}{\bf \Upsilon}_{j}{\bf \Upsilon}_{j}^H{\bf H}_{k}^H)}{n_c}},
\end{multline}
 where $\sigma^2$ is the noise variance, $n_c$ is the codeword length of the common part, $\epsilon_c$ is the maximum tolerable BER for the common part, and ${\bf S}_{ck}={\bf H}_{k}{\bf \Upsilon}{\bf \Upsilon}^H{\bf H}_{k}^H$ is the covariance matrix of the common part at the receiver of user $k$.  {Since all users have to detect ${\bf s}_{c}$, its transmission rate $r_{c}$ has to be supportable by all users.} Hence, $r_c$ has to obey
\begin{equation}\label{11}
    r_{c}\leq\min_k(r_{ck}).
\end{equation}

 After detecting ${\bf s}_{c}$, user ${k}$ removes ${\bf s}_{c}$ from its received signal and then detects ${\bf s}_{k}$. Thus, the maximum achievable rate for ${\bf s}_{k}$  is
\begin{multline}
       \label{12}
    r_{pk}=
\log \big|{\bf I} +\big(\!\sigma^2{\bf I}\!+\!\sum_{j\neq k}{\bf H}_{k}{\bf \Upsilon}_{j}{\bf \Upsilon}_{j}^H{\bf H}_{k}^H\big)^{-1}{\bf S}_{k} \big|
\\-Q^{-1}(\epsilon_p)\sqrt{\frac{2\text{Tr}({\bf S}_{k}(\sigma^2{\bf I}\!+\!\!\sum_j\!{\bf H}_{k}{\bf \Upsilon}_{j}{\bf \Upsilon}_{j}^H{\bf H}_{k}^H)^{-1})}{n_p}},
\end{multline}
where $n_p$ is the codeword length of the private message, $\epsilon_p$ is  the maximum tolerable BER for the private message, and ${\bf S}_{k} = {\bf H}_{k}{\bf \Upsilon}_{k}{\bf \Upsilon}_{k}^H{\bf H}_{k}^H$ is the covariance matrix of  the private message for user ${k}$ at its receiver.

The rate of user $k$ is 
    $r_{k}=z_{k}+r_{pk}$,
where $z_{k}$ is the rate portion of ${\bf s}_{c}$ dedicated to user $k$, which is an optimization variable \cite{mao2022rate}. The EE of user $k$ is
\begin{equation}
e_{k}=\frac{z_{k}+r_{pk}}{p_k(\{{\bf \Upsilon}\})},    
\end{equation}
where $p_k(\{{\bf \Upsilon}\})$ is the power consumption at supporting the rate $r_k$, which is given by \cite[Eq. (15)]{soleymani2024rate}
 \begin{equation}\label{(15)}
     p_{k}(\{{\bf \Upsilon}\})={P_s+\eta\text{Tr}\left({\bf \Upsilon}_{k}{\bf \Upsilon}_{k}^H\right)
     +\frac{\eta}{K} \text{Tr}\left({\bf \Upsilon}{\bf \Upsilon}^H\right)
     },
 \end{equation}
 where 
$\eta$ denotes the inverse power efficiency of the BS, and $P_s$ represents the system's static power consumption associated with data transmission to a user, as specified in \cite[Eq. (27)]{soleymani2022improper}. For notational simplicity, we omit the dependency of $p_{k}$ on the beamforming matrices in the expressions hereafter. Additionally, given the achievable rate $r_k$, the transmission delay of user $k$ for a message of length $l_k$ is 
$d_k={l_k}/{r_k}
$.    
Note that in the MIMO RS, a message with the length $l_k$ is split into two parts as described above. We assume that there are at least $l_k$ bits in the buffer of the BS awaiting transmission to user $k$.

\begin{remark}[ {Discussion on the reliability constraint}]
 {Reliability is explicitly accounted for through the FBL normal approximation, in which the decoding error probability is inherently embedded in the achievable rate expression \cite{polyanskiy2010channel}. In particular, the adopted rate formulation incorporates the target error probability $\epsilon$, ensuring that the resulting transmission rates satisfy the prescribed reliability requirement by design.}

 {Under the considered RSMA framework, successful decoding of the private stream requires prior decoding of the common stream followed by successive interference cancellation. Consequently, the overall decoding error probability at a user can be expressed as $\epsilon = \epsilon_c + (1-\epsilon_c)\epsilon_p$, which admits the conservative upper bound $\epsilon \leq \epsilon_c + \epsilon_p$. This bound provides a tractable and widely accepted characterization of reliability in RSMA systems operating under FBL constraints~\cite[Sec.~II.D]{soleymani2023optimization}.}
    
\end{remark}

\subsection{Problem Statement}
We study the latency-EE trade-off of RSMA by considering an OF, including both the minimum EE and maximum delay of users with appropriate weights. More specifically, we solve the following optimization problem
\begin{subequations}\label{17}
\begin{align}
\label{17a}
 \underset{\{{\bf \Upsilon}\},\bf{\Psi},\{{\bf z}\}
 }{\min}\,\,\,  & 
 \alpha\max_k\!\! \left(\frac{l_k}{r_{pk}+z_k}\right)\!\! -\!(1-\alpha)\!\min_k\!\! \left(\frac{r_{pk}+z_k}{p_k}\right)\!\! \\
  \text{s.t.}\,\,   \,
\label{17c}  & \text{Tr}({\bf \Upsilon}{\bf \Upsilon}^H+\sum_k{\bf \Upsilon}_k{\bf \Upsilon}_k^H)\leq P,
\\
\label{17d}
 &
z_{k}\geq 0,\,\,\,\,\forall k,
\\
\label{17e}
&
\sum_kz_{k}\leq \min_k(r_{ck}),
\\
\label{17f}
&
|\psi_m|\leq 1,\,\,\forall m,
\\
\label{17g}
&
\mathbf{H}_{k}\left({\bf \Psi}\right)= 
\mathbf{G}_{k}{\bf \Psi}\mathbf{G}+\mathbf{F}_{k}, \,\,\forall k,
 \end{align}
\end{subequations}
where $0\leq\alpha\leq 1$ (or $1-\alpha$) is the priority weight of latency (or EE), $P$ is the power budget of the BS,  \eqref{17c} is the power constraint,  \eqref{17d} ensures non-negative achievable rates, \eqref{17e} is the detectability constraint for ${\bf s}_c$, \eqref{17f} guarantees feasible values of the RIS coefficients, and \eqref{17g} accounts for the channel dependency on the RIS coefficients. 
The constraints in \eqref{17c}, \eqref{17d}, \eqref{17f}, and \eqref{17g} are convex. However, \eqref{17e} is non-convex since $r_{ck}$ is non-concave in $\{{\bf \Upsilon}\}$ and $\bf{\Psi}$. Additionally, the OF of \eqref{17} has a fractional structure, and it is neither a convex nor a concave function of the optimization variables. Therefore, \eqref{17} falls into the class of fractional matrix programming (FMP) optimization problems. 

\section{Proposed Solution}\label{sec:iii}
We derive a suboptimal solution for \eqref{17} by employing alternating optimization (AO) and the framework introduced in \cite{soleymani2025framework}. More specifically, the procedure is initialized with a feasible point $\{{\bf \Upsilon}^{(0)} \}$ and ${\bf \Psi}^{(0)}$, after which $\{{\bf \Upsilon} \}$ and ${\bf \Psi}$ are updated in an alternating manner. In the $i$-th iteration, we first update $\{{\bf \Upsilon}\}$ while keeping ${\bf \Psi}$ fixed at ${\bf \Psi}^{(i-1)}$. We then alternate and optimize ${\bf \Psi}$ when  $\{{\bf \Upsilon}\}$ is kept fixed at $\{{\bf \Upsilon}^{(i)}\}$. Updating $\{{\bf \Upsilon}\}$ (or ${\bf \Psi}$) is still challenging even when ${\bf \Psi}$ (or $\{{\bf \Upsilon}\}$) is kept fixed. Indeed, the rates are neither concave nor convex in $\{{\bf \Upsilon}\}$ or ${\bf \Psi}$. Additionally, the OF of \eqref{17} is a summation of fractional functions, which further complicates solving \eqref{17}. 
\subsection{Updating Beamforming Matrices}\label{sec:iii-a}

With ${\bf \Psi}$ held constant at ${\bf \Psi}^{(i-1)}$, \eqref{17} can be reformulated as 
\begin{subequations}\label{17-sur}
\begin{align}
 \underset{\{{\bf \Upsilon}\},\{{\bf z}\}
 }{\min}\!  & 
\left\{\! 
  \alpha\max_k d_k -(1-\alpha)\min_ke_k\!\right\}
  \\
  \text{s.t.}\  &
\eqref{17c}, \eqref{17d},\eqref{17e},
 \end{align}
\end{subequations}
which falls into FMP optimization problems.  To solve \eqref{17-sur}, we leverage the results in \cite[Lemma 1]{soleymani2025framework} and \cite[Lemma 2]{soleymani2025framework}. For this purpose, we need a concave lower bound for $r_{pk}$ and $r_{ck}$ for all $k$, presented in the following lemma. 

\begin{lemma}[\!\cite{soleymani2024rate}]\label{lem-4}
All feasible $\{{\bf \Upsilon}\}$ satisfy the 
inequalities
\begin{multline*}
r_{pk}\!\geq\! \tilde{r}_{pk}^{(i)}\! =\! 2\Re\!\left\{\!\text{{\em Tr}}\!\left(\!
{\bf A}_{k}{\bf \Upsilon}_{k}^H
\bar{\mathbf{H}}_{k}^H\!\right)\!\right\}
+\!2\!\sum_{j\neq k}\!\Re\!\left\{\!\text{{\em Tr}}\!\left(\!
{\bf A}_{kj}{\bf \Upsilon}_{j}^H
\bar{\mathbf{H}}_{k}^H\!\right)\!\right\}
\\
+a_{k}-
\text{{\em Tr}}\left(
{\bf B}_{k}\left(\sigma^2{\bf I}\!+\!\!\sum_j\!\bar{\bf H}_{k}{\bf \Upsilon}_{j}{\bf \Upsilon}_{j}^H\bar{\bf H}_{k}^H\right)
\right),
\end{multline*}
\begin{multline*}
r_{ck}\geq \tilde{r}_{ck}^{(i)} = a_{ck}
\\
+2\Re\left\{\text{{\em Tr}}\left(
{\bf A}_{ck}{\bf \Upsilon}^H
\bar{\mathbf{H}}_{k}^H\right)\right\}
+2\sum_{j}\Re\left\{\text{{\em Tr}}\left(
{\bf A}_{ckj}{\bf \Upsilon}_{j}^H
\bar{\mathbf{H}}_{k}^H\right)\right\}
\\
-
\text{{\em Tr}}\!\!\left(\!\!
{\bf B}_{ck}\!\!\left(\!\!\sigma^2{\bf I}\!+\bar{\bf H}_{k}{\bf \Upsilon}{\bf \Upsilon}^H\bar{\bf H}_{k}^H+\!\!\sum_j\!\bar{\bf H}_{k}{\bf \Upsilon}_{j}{\bf \Upsilon}_{j}^H\bar{\bf H}_{k}^H\!\!\right)\!\!
\right),\!\!
\end{multline*}
where we have:
\begin{align*}
{\bf C}_{k}&=\!\!\left(\!\!\sigma^2{\bf I}\!+\!\!\sum_{j\neq k}\bar{\bf T}_{kj}\!\!\right)^{-1}\!\!\bar{\bf T}_{kk},
\,\,
{\bf C}_{ck}=\!\!\left(\!\!\sigma^2{\bf I}\!+\!\!\sum_{j}\bar{\bf T}_{kj}\!\!\right)^{-1}\!\!\bar{\bf S}_{ck},
\\
e_{k}&=2\text{\em Tr}\left(\left(\sigma^2{\bf I}+\sum_{j}\bar{\bf T}_{kj}\right)^{-1}\bar{\bf T}_{kk}\right),
\\
e_{ck}&=2\text{\em Tr}\!\left(\!\!\left(\!\sigma^2{\bf I}+\bar{\bf S}_{ck}+\sum_{j}\bar{\bf T}_{kj}\right)^{-1}\bar{\bf S}_{ck}\!\right)\!,\!
\\
a_{k}&\!=\!\ln\left|{\bf I}+{\bf C}_{k}\right|
\!-\!
\text{{\em Tr}}\left(
{\bf C}_{k}
\right)\!
-\!\frac{Q^{-1}(\epsilon_p)}{2\sqrt{n}_p}\!\!
\left(\!\!\sqrt{e_k}\!+\!\frac{2N}{\sqrt{e_k}}\!
\right)\!\!,
\\
a_{ck}&=\!\ln\left|{\bf I}\!+\!{\bf C}_{ck}\right|
\!-\!
\text{{\em Tr}}\!\left(
{\bf C}_{ck}
\right)\!
-\!\frac{Q^{-1}(\epsilon_c)}{2\sqrt{n}_c}\!\!
\left(\!\!\!\sqrt{e_{ck}}\!+\!\frac{2N}{\sqrt{e_{ck}}}\!
\right)\!\!,
\\
{\bf A}_{k}&\!=\!\!\left(\!\!\sigma^2{\bf I}\!+\!\sum_{j\neq k}\bar{\bf T}_{kj}\!\!\right)^{-1}\!\!\!\!\bar{\mathbf{H}}_{k}\bar{{\bf \Upsilon}}_{k}\!,
\\
{\bf A}_{ck}&\!=\!\!\left(\!\!\sigma^2{\bf I}\!+\!\sum_{j}\bar{\bf T}_{kj}\!\!\right)^{-1}\!\!\!\!\bar{\mathbf{H}}_{k}\bar{{\bf \Upsilon}}\!,
\\
{\bf A}_{kj}\!&\!=
\frac{Q^{-1}(\epsilon_p)}{\sqrt{n_pe_k}}
\left(\sigma^2{\bf I}+\sum_{j}\bar{\bf T}_{kj}\right)^{-1}\bar{\mathbf{H}}_{k}\bar{{\bf \Upsilon}}_{j},
\\
{\bf A}_{ckj}\!&\!=\frac{Q^{-1}(\epsilon_c)}{\sqrt{n_ce_{ck}}}\!\!
\left(\!\!\sigma^2{\bf I}\!+\!\bar{\bf S}_{ck}+\!\! \sum_{j}\bar{\bf T}_{kj}\!\!\right)^{-1}\!\!\bar{\mathbf{H}}_{k}\bar{{\bf \Upsilon}}_j\!,
\\
{\bf B}_{k}&\!=\!\!\left(\!\!\sigma^2{\bf I}+\sum_{j\neq k}\bar{\bf T}_{kj}\!\!\right)^{-1}-\left(\!\!\sigma^2{\bf I}+\sum_{j}\bar{\bf T}_{kj}\!\!\right)^{-1}
+\!
\frac{Q^{-1}(\epsilon_p)}{\sqrt{n_pe_k}}
\\&\!\!
\times\!\!\left(\!\!\sigma^2{\bf I}\!+\!\!\sum_{j}\bar{\bf T}_{kj}\!\!\right)^{-1}
\left(\!\!\sigma^2{\bf I}\!+\!\!\sum_{j\neq k}\bar{\bf T}_{kj}\!\!\right)
\left(\!\!\sigma^2{\bf I}\!+\!\!\sum_{j}\bar{\bf T}_{kj}\!\!\right)^{-1},
\\
{\bf B}_{ck}&=\left(\sigma^2{\bf I}+\sum_{j}\bar{\bf T}_{kj}\right)^{-1}-\left(\sigma^2{\bf I}+\bar{\bf S}_{ck}+ \sum_{j}\bar{\bf T}_{kj}\right)^{-1}
\\&
\ \ \ +
\frac{Q^{-1}(\epsilon_c)}{\sqrt{n_ce_k}}
\left(\sigma^2{\bf I}+\bar{\bf S}_{ck}+\sum_{j}\bar{\bf T}_{kj}\right)^{-1}
\\&
\ \ \  \times
\left(\sigma^2{\bf I}+\sum_{j}\bar{\bf T}_{kj}\right)
\left(\sigma^2{\bf I}+\bar{\bf S}_{ck}+\sum_{j}\bar{\bf T}_{kj}\right)^{-1},
\end{align*}
where $\bar{\bf T}_{kj}=\bar{\bf H}_{k}\bar{\bf \Upsilon}_{j}\bar{\bf \Upsilon}_{j}^H\bar{\bf H}_{k}^H$, 
$\bar{\bf S}_{ck}=\bar{\bf H}_k\bar{\bf \Upsilon}\bar{\bf \Upsilon}^H\bar{\bf H}_k^H $, 
$\bar{\bf H}_k={\bf H}_k({\bf \Psi}^{(i-1)} )$, $\bar{\bf \Upsilon}={\bf \Upsilon}^{(i-1)}$, and $\bar{\bf \Upsilon}_j={\bf \Upsilon}_j^{(i-1)}$ for all $k,j$.
\end{lemma} 
Upon leveraging \cite[Lemma 1]{soleymani2025framework}, \cite[Lemma 2]{soleymani2025framework}, and the concave bounds in Lemma \ref{lem-4}, \eqref{17-sur} is  rewritten as 
\begin{subequations}\label{17-sur2}
\begin{align}
 \underset{\{{\bf \Upsilon}\},\{{\bf z}\},\{{\bf t}\},\{{\bf u}\},e,d
 }{\min}\,\,\,  & 
  \alpha d-(1-\alpha)e  
  \\
  \text{s.t.}\,\,   \,\,\,   \,
& \eqref{17c}, \eqref{17d},
\\
&
\sum_kz_{k}\leq \min_k(\tilde{r}_{ck}^{(i)}),
\\
\label{12c}&
\tilde{r}_{pk}^{(i)}+z_k\geq t_k\geq 0, \,\,\, \forall k,
\\
\label{12d}&
\tilde{r}_{pk}^{(i)}+z_k\geq u_k^2, \,\,\,u_k\geq 0,\,\,\, \forall k,
\\
&
\frac{l_k}{t_k}\leq d, \,\,\, \forall k,
\\
&
2\lambda_k^{(i)} u_k-\left(\lambda_k^{(i)}\right)^2p_k\geq e,\,\,\, \forall k,
 \end{align}
\end{subequations}
where $\{{\bf t}\}=\{t_1,t_2,\cdots,t_K\}$, $\{{\bf u}\}=\{u_1,u_2,\cdots,u_K\}$, $e$, and $d$  are auxiliary variables, and $\lambda_k^{(i)}$ is constant, given by
\begin{equation}
    \lambda_k^{(i)}=\frac{\sqrt{r_k(\{{\bf \Upsilon}^{(i-1)}\},{\bf \Psi}^{(i-1)})}}{p_k(\{{\bf \Upsilon}^{(i-1)}\})}
\end{equation}
for all $k$. Note that the constraints of \eqref{17-sur2} are convex, and its OF is linear, making \eqref{17-sur2} a convex optimization problem. 
Solving \eqref{17-sur2} yields $\{{\bf \Upsilon}^{(i)}\}$.
\subsection{Updating RIS Scattering Matrix}
When $\{{\bf \Upsilon}\}$ is kept fixed at $\{{\bf \Upsilon}^{(i)}\}$, \eqref{17} reduces to
\begin{subequations}\label{17-sur-t}
\begin{align}
 \underset{{\bf \Psi},\{{\bf z}\}
 }{\min} &\left\{\! 
  \alpha\max_kd_k -(1-\alpha)\min_k e_k\!\right\} \\
  \text{s.t.}\,\,   &
 \eqref{17d},\eqref{17e},\eqref{17f},\eqref{17g}.
 \end{align}
\end{subequations}
When beamforming matrices are kept fixed, $e_k$ becomes a linear function of $r_k$, making the solution of \eqref{17-sur-t} simpler than that of \eqref{17-sur}. Nevertheless,  \eqref{17-sur-t} is still an FMP optimization problem, and we leverage \cite[Lemma 1]{soleymani2025framework} and a concave lower bound for $r_{pk}$ and $r_{ck}$ for all $k$ to solve it. The lemma below represents a concave lower bound for $r_{pk}$ and $r_{ck}$.
\begin{lemma}[\!\cite{soleymani2024rate}]\label{lem-2}
All feasible ${\bf \Psi}$ satisfy the inequalities
\begin{multline}
r_{pk}\!\geq\! \hat{r}_{pk}^{(i)}\! =\! 2\Re\!\left\{\!\text{{\em Tr}}\!\left(\!
{\bf A}_{k}\bar{\bf \Upsilon}_{k}^H
{\mathbf{H}}_{k}^H\!\right)\!\right\}
+\!2\!\sum_{j\neq k}\!\Re\!\left\{\!\text{{\em Tr}}\!\left(\!
{\bf A}_{kj}\bar{\bf \Upsilon}_{j}^H
{\mathbf{H}}_{k}^H\!\right)\!\right\}
\\
+a_{k}-
\text{{\em Tr}}\left(
{\bf B}_{k}\left(\sigma^2{\bf I}\!+\!\!\sum_j\!{\bf H}_{k}\bar{\bf \Upsilon}_{j}\bar{\bf \Upsilon}_{j}^H{\bf H}_{k}^H\right)
\right),
\end{multline}
\begin{multline}
r_{ck}\geq \hat{r}_{ck}^{(i)} = a_{ck}
\\
+2\Re\left\{\text{{\em Tr}}\left(
{\bf A}_{ck}\bar{\bf \Upsilon}^H
{\mathbf{H}}_{k}^H\right)\right\}
+2\sum_{j}\Re\left\{\text{{\em Tr}}\left(
{\bf A}_{ckj}\bar{\bf \Upsilon}_{j}^H
{\mathbf{H}}_{k}^H\right)\right\}
\\
-
\text{{\em Tr}}\!\!\left(\!\!
{\bf B}_{ck}\!\!\left(\!\!\sigma^2{\bf I}\!+{\bf H}_{k}\bar{\bf \Upsilon}\bar{\bf \Upsilon}^H{\bf H}_{k}^H+\!\!\sum_j\!{\bf H}_{k}\bar{\bf \Upsilon}_{j}\bar{\bf \Upsilon}_{j}^H{\bf H}_{k}^H\!\!\right)\!\!
\right)\!\!,\!\!
\end{multline}
where all coefficients are defined as in Lemma \ref{lem-4} by setting $\bar{\bf \Upsilon}={\bf \Upsilon}^{(i)}$, and $\bar{\bf \Upsilon}_j={\bf \Upsilon}_j^{(i)}$, $\forall j$. 
\end{lemma}
Upon utilizing \cite[Lemma 1]{soleymani2025framework} and Lemma \ref{lem-2}, \eqref{17-sur-t} is rewritten as
\begin{subequations}\label{17-sur-t2}
\begin{align}
 \underset{{\bf \Psi},\{{\bf z}\},\{{\bf t}\},e
 }{\min}\,\,\,  & 
  \alpha\max_k\left\{ \frac{l_k}{t_k}\right\}-(1-\alpha)e \\
  \text{s.t.}\,\,   \,
& \eqref{17d},\eqref{17f},\eqref{17g},
\,\,\sum_kz_{k}\leq \min_k(\hat{r}_{ck}^{(i)}),
\\
\label{14c}&
\hat{r}_{pk}^{(i)}\!+\!z_k\!\geq\! \max\{ t_k,ep_k(\{{\bf \Upsilon}^{(i)} \})\}\!\geq 0, \,\, \forall k,\!\!
 \end{align}
\end{subequations}
where $\{{\bf t}\}=\{t_1,t_2,\cdots,t_K\}$ and $e$ are auxiliary variables. The solution of \eqref{17-sur-t2} gives ${\bf \Psi}^{(i)}$.

 The convergence behavior of the proposed algorithm follows from the monotonic reduction of the objective function achieved in each iteration. Given that the objective of \eqref{17} is bounded from below, convergence of the iterative sequence is ensured. Additionally, the surrogate functions comply with the established majorization minimization (MM) principles, including exactness at the operating point, gradient consistency, and global bounding \cite{soleymani2025framework, sun2017majorization, soleymani2020improper}. These properties collectively guarantee convergence to a stationary solution of \eqref{17}. A summary of the solution is provided in Algorithm I, where $\beta_k$ is the objective function of \eqref{17}. 
\doublespacing 
\begin{table}[htb]
\begin{tabular}{l}
\hline 
 \textbf{Algorithm I}:  {Summary of the proposed solution.}  \\
\hline
\hspace{0.1cm}{\textbf{Initialization}}:
Set $\delta$,  $i=1$,  $\{\mathbf{W}\}=\{\mathbf{W}^{(0)}\}$, ${\bf \Psi}={\bf \Psi}^{(0)}$ \\
\hline 
\hspace{0.1cm}
\textbf{While} $\big(\underset{\forall k}{\max}\,\beta^{(i)}_{k}-\underset{\forall k}{\max}\,\beta^{(i-1)}_{k}\big)/\underset{\forall k}{\max}\,\beta^{(i-1)}_{k}\geq\delta$\\ 
\hspace{.6cm}{Calculate $\tilde{r}_{pk}^{(i)}$ and $\tilde{r}_{ck}^{(i)}$ according to Lemma \ref{lem-4}}\\ 
\hspace{.6cm}{Solve \eqref{17-sur2} to find $\{{\bf W}^{(i)} \}$}\\
\hspace{.6cm}{Calculate $\hat{r}_{pk}^{(i)}$ and $\hat{r}_{ck}^{(i)}$ according to Lemma \ref{lem-2}}\\ 
\hspace{.6cm}{Solve \eqref{17-sur-t2} to find ${\bf \Psi}^{(i)} $}\\
\hspace{.6cm}$i=i+1$\\
\hspace{0.1cm}\textbf{End (While)}, 
 {{\bf Return} $\{\mathbf{W}^{(\star)}\}$ and ${\bf \Phi}^{(\star)}$}\\
\hline 
\end{tabular}  
\end{table}
\singlespacing\normalsize

\subsection{ {Computational Complexity Analysis}}
 This subsection provides an order-level characterization of the computational complexity of the proposed algorithm in terms of the number of multiplications. As described in Algorithm~I, the solution is obtained via an iterative AO framework combined with an MM procedure. In each iteration, two convex optimization problems are solved sequentially: one corresponding to the update of the beamforming matrices $\{{\boldsymbol{\Upsilon}}\}$ and the other one to the update of the RIS scattering matrix ${\boldsymbol{\Psi}}$. Consequently, the overall complexity is determined by the sum of the complexities associated with solving these two convex subproblems.

 {Since both optimization problems are convex, they are solved using standard interior-point methods (IPMs). According to convex optimization theory, the number of Newton steps required by IPMs scales proportionally to the square root of the total number of constraints \cite[Chapter~11]{boyd2004convex}. Therefore, the iteration complexity primarily depends on the constraint set size of each subproblem.}

\subsubsection{Beamforming Update}
Updating $\{{\boldsymbol{\Upsilon}}\}$ requires solving \eqref{17-sur2}, which contains $6K+1$ constraints. For order analysis, this can be approximated as $6K$, implying that the number of Newton iterations scales as $\mathcal{O}(\sqrt{6K})$. The dominant computational cost per Newton step arises from evaluating the surrogate rate expressions for all users. Under the RSMA transmission strategy considered, two rate terms have to be computed per user, corresponding to the common and private messages. Since both rates exhibit comparable computational complexity, we conservatively characterize the per-user cost using the common-message rate, which involves slightly more operations.

Although the achievable rates are originally expressed using log-det functions involving SINR matrices, the proposed MM framework employs the lower bounds derived in Lemma \ref{lem-4}. These surrogate functions are quadratic in $\{{\boldsymbol{\Upsilon}}\}$, with coefficients that remain fixed throughout the IPM iterations. As both the coefficients and channel matrices are computed once prior to the Newton updates, their contribution is negligible compared to the repeated matrix multiplications required within each Newton step. Specifically, the dominant term stems from evaluating expressions of the form
\[
\mathrm{Tr}\!\left(
{\bf B}_{ck}\!\left(\sigma^2{\bf I}
+ \bar{\bf H}_{k}{\boldsymbol{\Upsilon}}{\boldsymbol{\Upsilon}}^H\bar{\bf H}_{k}^H
+ \sum_j \bar{\bf H}_{k}{\boldsymbol{\Upsilon}}_{j}{\boldsymbol{\Upsilon}}_{j}^H\bar{\bf H}_{k}^H
\right)
\right),
\]
which requires computing matrix products such as
$\bar{\bf H}_{k}{\boldsymbol{\Upsilon}}{\boldsymbol{\Upsilon}}^H\bar{\bf H}_{k}^H$
for all $K$ users. Recall that $\bar{\bf H}_{k} \in \mathbb{C}^{N_u \times N_{BS}}$ and 
${\boldsymbol{\Upsilon}} \in \mathbb{C}^{N_{BS} \times N}$, where 
$N = \min(N_{BS}, N_u)$. For worst-case scaling, $N$ can be approximated by $N_{BS}$. The required matrix multiplications therefore scale as
$\mathcal{O}\!\left(N_{BS}^2 (2N_{BS} + N_u)\right)$. 
Since these operations are performed across $K$ users, the dominant complexity per Newton step becomes
$\mathcal{O}\!\left(K^2 N_{BS}^2 (2N_{BS} + N_u)\right)$.
Combining this with the IPM iteration count yields the overall beamforming update complexity:
\begin{equation}
\mathcal{O}\!\left(K^2 N_{BS}^2 \sqrt{6K} (2N_{BS} + N_u)\right).
\end{equation}
Importantly, this complexity term is independent of the RIS size $M$. This follows from the fact that the channel matrices are computed once prior to solving \eqref{17-sur2} and remain constant throughout the IPM iterations.

\subsubsection{RIS Scattering Matrix Update}
Updating ${\boldsymbol{\Psi}}$ requires solving \eqref{17-sur-t2}, which contains $3K+M+1$ constraints. In compact matrix form, the unit-modulus constraints can be represented efficiently; however, to derive an upper bound, we conservatively consider $M$ scalar constraints. Consequently, the Newton iteration count scales as $\mathcal{O}(\sqrt{3K+M})$.
While the surrogate rate evaluations remain of comparable order to the beamforming update, this step additionally requires recomputation of the effective channel matrices, since they explicitly depend on the RIS coefficients. At each Newton step, the equivalent channel matrices must therefore be updated for all $K$ users.
The dominant operation in updating the channels is the evaluation of the cascaded channel term ${\bf G}_k\boldsymbol{\Psi}{\bf G}$, where $\boldsymbol{\Psi}$ is diagonal. This operation requires approximately $\mathcal{O}(N_{BS} M N_u)$
multiplications. Accounting for both rate evaluations and channel updates, the RIS update complexity per AO iteration becomes
\begin{multline}
\mathcal{O}\big(
K^2 N_{BS}^2 \sqrt{3K+M}(2N_{BS}+N_u)
\\+ K M N_u N_{BS} \sqrt{3K+M}
\big).
\end{multline}

\subsubsection{Overall Complexity}

Combining the beamforming and RIS update costs over $N_{\max}$ AO iterations, the total computational complexity is on the order of
\begin{multline}
\mathcal{O}\big(
N_{\max} K^2 N_{BS}^2 (\sqrt{6K}+\sqrt{3K+M})(2N_{BS}+N_u) \\
+ N_{\max} K M N_u N_{BS} \sqrt{3K+M}
\big).
\end{multline}
In summary, the computational complexity increases with $K^{5/2}$, $N_{BS}^3$, $N_u$, and $M^{3/2}$. The dependence on $M^{3/2}$ characterizes the marginal impact of RIS size under fixed $(K, N_{BS}, N_u)$, whereas the full scaling behavior is governed by the expression above when multiple system dimensions vary jointly.

\subsubsection{Comparison with SDMA}
 The computational complexity of SDMA is in the same order as that of RSMA for the considered setup. Indeed, to obtain an RSMA solution, one needs to compute $2K$ rates in total, i.e., two rates per user: one for the common message and one for the private message. By contrast, for deriving a solution for SDMA, only one rate per user has to be calculated. However, the computational complexity of computing each rate in both cases is in the same order. Therefore, the computational complexity of SDMA scales with $K^{5/2}$, $N_{BS}^{3}$, $N_u$, and $M^{3/2}$, which yields a similar complexity order as RSMA.

\section{Numerical Results}\label{sec-iv}
\begin{figure}[t]
    \centering
           \includegraphics[width=.46\textwidth]{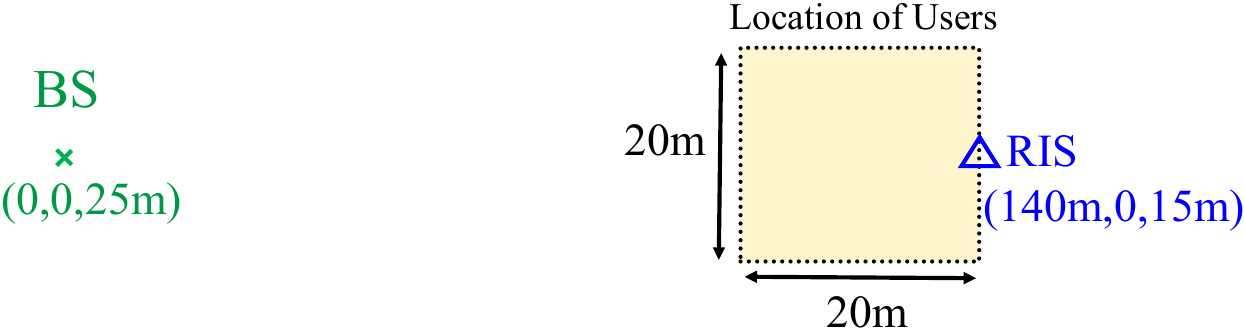}
    \caption{ {Network topology for numerical evaluations.}}
	\label{Fig-sim-topo}  
\end{figure}

This section presents numerical results obtained through Monte Carlo simulations.  {The network topology for the numerical evaluations is depicted in Fig. \ref{Fig-sim-topo}.} Moreover, the simulation parameters and setup are based on \cite[Appendix G]{soleymani2025framework}. 
Additionally, we set $M=20$, $\epsilon=2\epsilon_p=2\epsilon_c=10^{-5}$, and $n_p=n_c=n=256$ bits.  {The performance of RSMA is benchmarked against a conventional SDMA baseline, obtained by setting the power of the common message to zero, i.e., ${\bf \Upsilon}=0 $.} In particular, legend definitions for the figures are provided below:
\begin{itemize}
\item {\bf RIS-RSMA} (or {\bf RIS-SDMA}): RSMA (or SDMA) transmission strategy tailored to a MIMO FBL system assisted by an RIS.

\item {\bf No-RIS-RSMA} (or {\bf No-RIS-SDMA}): RSMA (or SDMA) framework designed for MIMO FBL scenarios operating without RIS support.

\item {\bf RIS-Rand-RSMA} (or {\bf RIS-Rand-SDMA}): RSMA (or SDMA) approach for RIS-aided MIMO FBL systems in which the RIS phase shifts are randomly generated with unit-modulus constraints.
\end{itemize}
 {The benchmarks selected are designed to isolate the individual and joint contributions of RIS deployment and RSMA-based interference management. Specifically, comparisons between RIS-assisted and No-RIS schemes quantify the propagation gains introduced by the RIS, whereas comparisons between RSMA and SDMA reveal the benefits of rate splitting under identical system assumptions. Moreover, the RIS-Rand configurations provide a reference illustrating the performance attainable without phase-shift optimization. This benchmark structure is particularly relevant given the absence of prior studies jointly examining RIS-assisted RSMA systems under FBL constraints with explicit EE-latency trade-off optimization.}

\begin{figure}[t]
    \centering
    \begin{subfigure}[t]{0.45\textwidth}
        \centering
           \includegraphics[width=.8\textwidth]{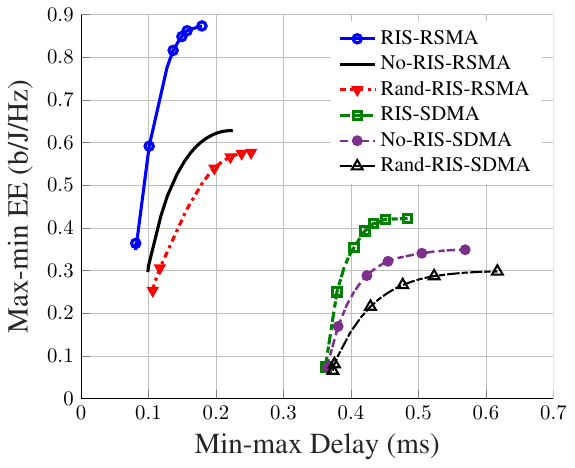}
        \caption{$P=10$ dB.}
    \end{subfigure}
    \begin{subfigure}[t]{0.45\textwidth}
        \centering
           \includegraphics[width=.8\textwidth]{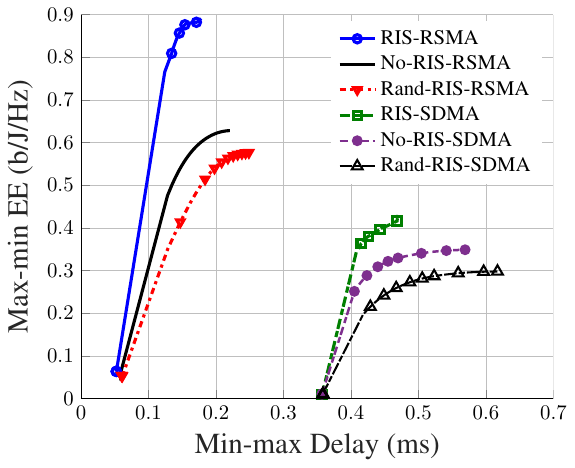}
        \caption{$P=20$ dB.}
    \end{subfigure}
    \caption{Latency-EE region ($N_{{BS}}=N_{u}=3$, and $K=4$).}
	\label{Fig-latency-energy}  
\end{figure}
\begin{figure}[t]
    \centering
    \begin{subfigure}[t]{0.24\textwidth}
        \centering
           \includegraphics[width=\textwidth]{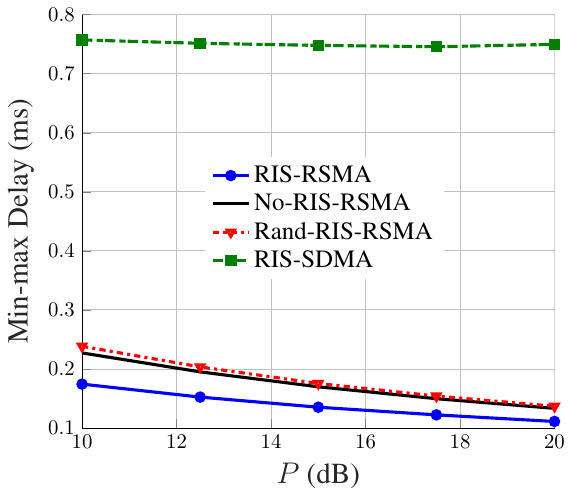}
        \caption{$K=4$.}
    \end{subfigure}
    \begin{subfigure}[t]{0.24\textwidth}
        \centering
       \includegraphics[width=\textwidth]{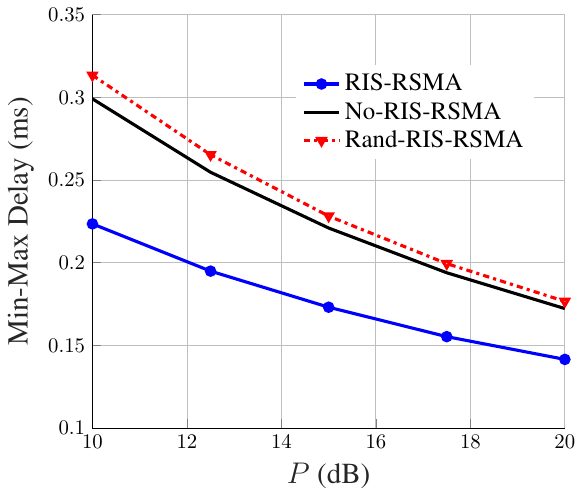}
        \caption{$K=5$.}
    \end{subfigure}%
    \caption{Average min-max delay versus $P$ ($N_{{BS}}=N_{u}=2$).}
	\label{Fig-sum-d}  
\end{figure}

Fig. \ref{Fig-latency-energy} demonstrates the trade-off between latency and EE, where the $x$-axis represents the min-max delay and the $y$-axis shows the max-min EE. Each RSMA scheme significantly reduces latency, while simultaneously enhancing the EE. As expected, reducing latency requires higher transmission power, which in turn erodes the EE. However, the extent of EE degradation is surprisingly severe, especially when latency has a higher priority than EE in the OF of \eqref{17}. In Fig. \ref{Fig-latency-energy}, we depict the latency-EE region for two values of $P$, i.e., $P=10$ dB and $P=20$ dB. While the maximum EE remains constant across both cases, the minimum achievable delay improves substantially, as $P$ increases. Indeed, to maximize EE, the BS has to transmit at a power lower than the power budget, when $P$ is sufficiently high \cite{buzzi2016survey}. By contrast, latency is a strictly decreasing function of $P$, as shown in Fig. \ref{Fig-sum-d}. Another key observation in Fig. \ref{Fig-latency-energy} concerns the gains of employing RIS. Although RIS offers modest EE improvements, its impact on latency is limited. 
 {Furthermore, RIS with random coefficients performs worse than the no-RIS baseline, highlighting the importance of adequate RIS optimization. Indeed, it is important to emphasize that random RIS phase shifts do not guarantee constructive combining. In multi-user interference-limited regimes, the reflected links may alter the interference structure, potentially leading to marginal or even negative performance variations relative to the No-RIS case for specific channel realizations.}

Fig. \ref{Fig-sum-d} illustrates the average maximum delay versus $P$, demonstrating that RSMA reduces the average maximum delay by an order of magnitude compared to SDMA. Specifically, the maximum delay achieved by SDMA in Fig. \ref{Fig-sum-d}a is at least 3.3 times (and up to 5.8 times) higher than that of RSMA. Equivalently, RSMA reduces  the average maximum delay to $23\%$ ($15\%$) compared to that of SDMA when $P=10$ dB ($P=20$ dB), as seen in Fig. \ref{Fig-sum-d}a. The average maximum delay also decreases as the BS power budget increases. Indeed, higher transmission power enables higher data rates, resulting in lower latency. Additionally, the average maximum delay rises significantly as $K$ increases from $4$ in Fig. \ref{Fig-sum-d}a to $5$ in Fig. \ref{Fig-sum-d}b. An RIS can notably reduce latency, but only when optimized. Overall, the most substantial performance gains are due to RSMA, with RIS offering some further, less pronounced, improvements.

\begin{figure}[t]
    \centering
    \begin{subfigure}[t]{0.24\textwidth}
        \centering
           \includegraphics[width=\textwidth]{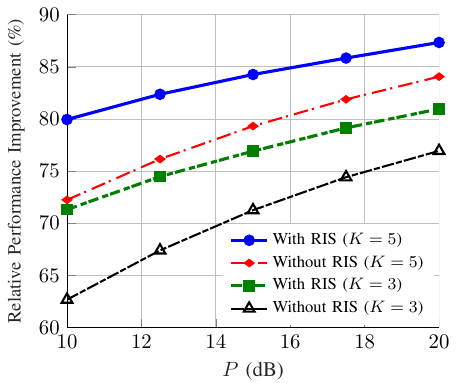}
        \caption{RSMA gains over SDMA.}
    \end{subfigure}
    \begin{subfigure}[t]{0.24\textwidth}
        \centering
       \includegraphics[width=\textwidth]{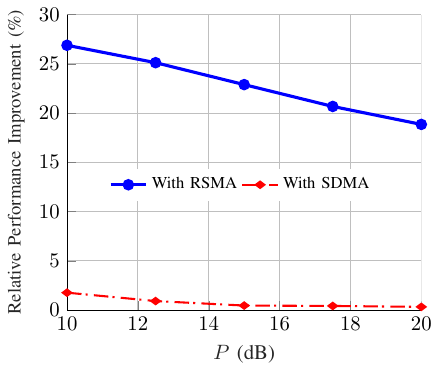}
        \caption{RIS gains over No-RIS ($K=4$).}
    \end{subfigure}%
    \caption{Delay reductions by employing RSMA and RIS versus $P$ ($N_{{BS}}=N_{u}=2$).}
	\label{Fig-2}  
\end{figure}
  {Fig. \ref{Fig-2} depicts the gains achieved by using RSMA, compared to SDMA, and RIS, compared to No-RIS. In Fig. \ref{Fig-2}a, RSMA substantially reduces the maximum delay, namely down to 13\% of that of SDMA. Furthermore, its gains become more pronounced as the number of users or the BS power budget increases. This is because higher values of $K$ or $P$ lead to more severe interference, thereby increasing the improvements of RSMA as an interference-management technique. Moreover, RIS further enhances the gains of RSMA by improving the quality of the effective channels, which has a similar impact to that of increasing the transmit power. Fig. \ref{Fig-2}b shows that while a RIS improves the channel quality and coverage, its stand-alone impact is marginal, particularly at higher transmit powers. This is because a RIS cannot effectively manage interference on its own in downlink scenarios with high user load. Additionally, as $P$ increases, the marginal benefits of RIS diminish due to its limited contribution to further channel enhancement. The considered setup is interference-limited ($K > N_{\text{BS}}$), where RSMA and RIS serve \textit{complementary roles}: RIS improves link quality, while RSMA manages the resulting interference. Their combination leads to a synergistic effect that significantly improves overall performance.}

\begin{figure}[t]
    \centering
    \begin{subfigure}[t]{0.25\textwidth}
        \centering
           \includegraphics[width=\textwidth]{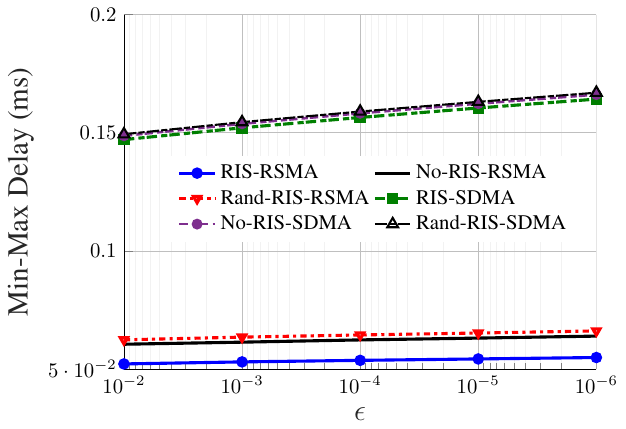}
        \caption{Average max delay.}
    \end{subfigure}
    \begin{subfigure}[t]{0.224\textwidth}
        \centering
           \includegraphics[width=\textwidth]{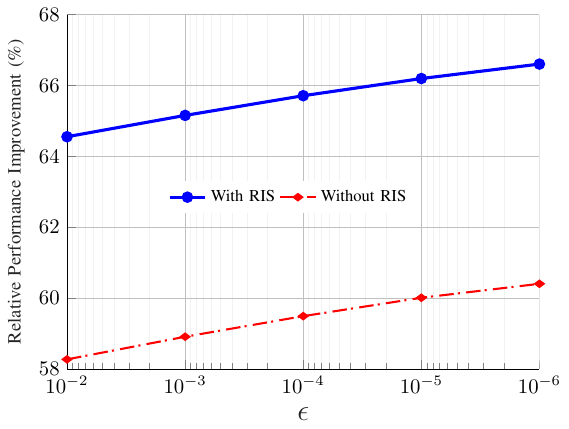}
        \caption{RSMA gains over SDMA.}
    \end{subfigure}
    \caption{Average min-max delay versus $\epsilon$ ($P=10$ dB, $N_{{BS}}=N_{u}=3$, and $K=4$).}
	\label{Fig-4}  
\end{figure}
Fig. \ref{Fig-4} shows the average maximum delay versus $\epsilon$. As depicted, tightening the reliability constraint leads to increased latency, since smaller values of $\epsilon$ impose more conservative transmission rates. Again, the amalgam of RSMA and RIS substantially outperforms the other schemes. Additionally, RSMA and RIS become synergetic, and their benefits increase when more reliable communication is needed. Finally, RSMA provides more significant gains than RIS.

\begin{figure}[t]
    \centering
           \includegraphics[width=.34\textwidth]{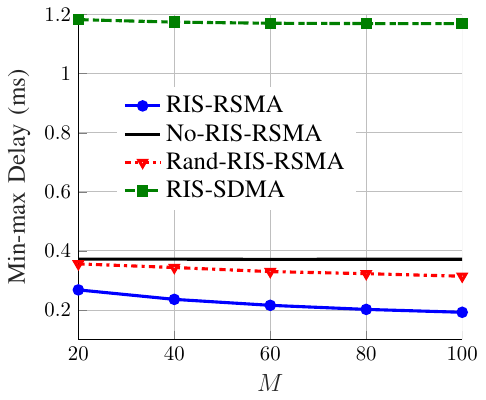}
    \caption{ {Average min-max delay versus $M$ ($P=10$ dB, $N_{{BS}}=N_{u}=2$, and $K=4$).}}
	\label{Fig-6} 
\end{figure}
 Fig. \ref{Fig-6} illustrates the impact of $M$ on the min-max delay. As the RIS size grows, all the schemes experience latency reduction; however, the improvement achieved by RSMA is more substantial. This stems from the interference-limited nature of this setup. Since $K>N$, RIS with SDMA cannot effectively mitigate interference, whereas RSMA is specifically designed to manage it. Thus, enlarging the RIS provides RSMA with stronger effective channels and higher SINR at the receivers, thereby amplifying its latency gains.  {The figure also indicates that random RIS configurations benefit from larger element counts, although their performance remains notably inferior to optimized RIS designs, highlighting the necessity of careful RIS optimization to fully realize the available gains.}

\begin{figure}[t]
    \centering
    \begin{subfigure}[t]{0.24\textwidth}
        \centering
           \includegraphics[width=\textwidth]{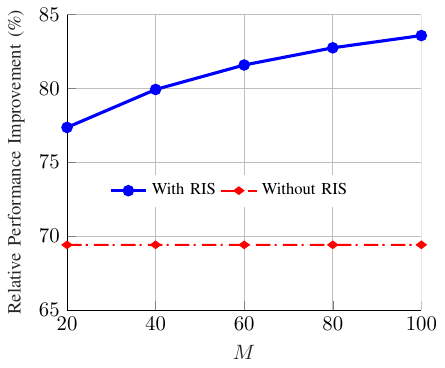}
        \caption{RSMA latency gains over SDMA.}
    \end{subfigure}
    \begin{subfigure}[t]{0.24\textwidth}
        \centering
           \includegraphics[width=\textwidth]{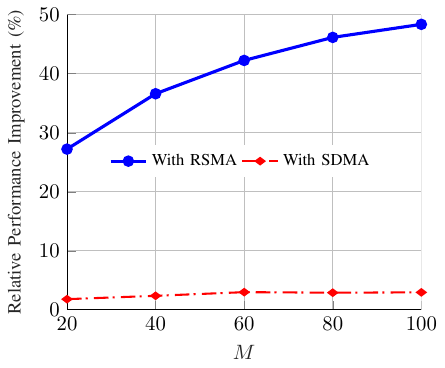}
        \caption{RIS latency gains over No-RIS.}
    \end{subfigure}
    \caption{ {RSMA and RIS latency gains versus $M$ ($P=10$ dB, $N_{{BS}}=N_{u}=2$, and $K=4$).}}
	\label{Fig-7} 
\end{figure}
 Fig. \ref{Fig-7} further examines the joint behavior of RSMA and RIS in two subfigures. Fig. \ref{Fig-7}a presents the latency reductions achieved by RSMA with and without RIS support. The results confirm the synergistic effect previously observed. When interference limits performance, RIS strengthens the effective channels, enabling RSMA to perform interference management more efficiently. Importantly, this synergy becomes increasingly prominent as $M$ grows. Fig. \ref{Fig-7}b depicts the benefits provided by the RIS when operating with RSMA and SDMA. Consistent with Fig. \ref{Fig-7}a, the RIS yields considerably larger gains when paired with RSMA, and these gains scale with $M$. In contrast, RIS improvements remain modest under SDMA, typically below 5\%, and increasing the RIS size does little to alter this outcome, reinforcing that RIS alone cannot resolve interference in overloaded settings.

\begin{figure}[t]
    \centering
           \includegraphics[width=.34\textwidth]{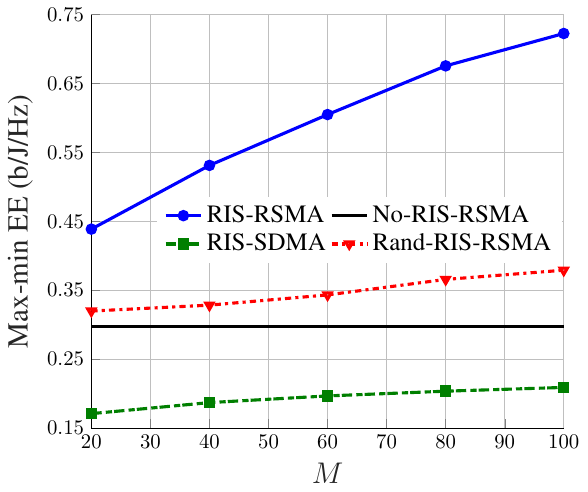}
    \caption{ {Average max-min EE versus $M$ ($P=10$ dB, $N_{{BS}}=N_{u}=3$, $P_c=3$ W, and $K=4$).}}
	\label{Fig-8}  
\end{figure}
Fig. \ref{Fig-8} illustrates the average max-min EE as a function of $M$. As with previous analyses, RSMA consistently outperforms SDMA and demonstrates improved efficiency as $M$ increases. The underlying trade-off remains intact: enlarging the RIS enhances the quality of the effective channels, which benefits both multiple-access schemes, yet RSMA unlocks significantly greater gains because it can exploit the higher effective SINR to manage interference more efficiently. The figure also highlights that enhancing channel quality via RIS alone offers limited EE gains under SDMA, further emphasizing that EE improvements in interference-limited regimes rely critically on the integration of an interference-resilient transmission strategy.

\begin{figure}[t]
    \centering
    \begin{subfigure}[t]{0.24\textwidth}
        \centering
           \includegraphics[width=\textwidth]{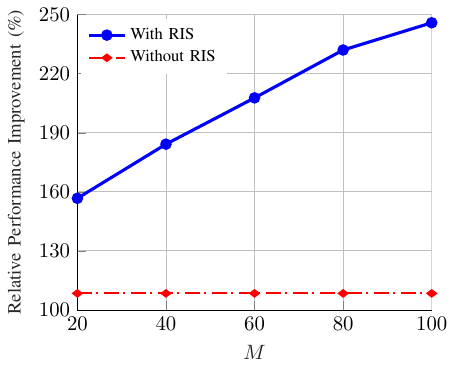}
        \caption{RSMA EE gains over SDMA.}
    \end{subfigure}
    \begin{subfigure}[t]{0.24\textwidth}
        \centering
           \includegraphics[width=\textwidth]{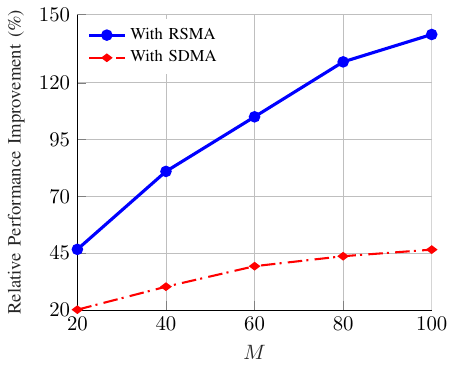}
        \caption{RIS EE gains over No-RIS.}
    \end{subfigure}
    \caption{RSMA and RIS EE gains versus $M$ ($P=10$ dB, $N_{{BS}}=N_{u}=3$, $P_c=3$ W, and $K=4$).}
	\label{Fig-9}  
\end{figure}
 Fig. \ref{Fig-9} examines the EE benefits associated with RSMA and RIS through two subfigures mirroring the structure of Fig. \ref{Fig-7}. Fig. \ref{Fig-9}a shows that RSMA achieves pronounced EE improvements, and the addition of an RIS further magnifies these gains as its size increases. This again reflects the complementary roles of the two technologies: RIS enhances channel quality, while RSMA converts these stronger links into effective interference mitigation. Fig. \ref{Fig-9}b focuses on the gains offered by the RIS when paired with RSMA or SDMA. Although RIS alone improves EE, its benefits remain significantly higher when used jointly with RSMA, particularly for larger $M$. Under SDMA, EE improvements by RIS grow only marginally with $M$, confirming that effective interference management is essential for unlocking the substantial EE potential of large RIS deployments.

 \begin{figure}[t]
    \centering
           \includegraphics[width=.34\textwidth]{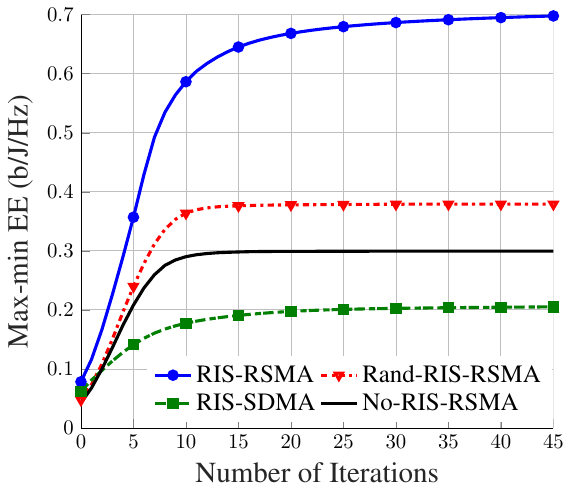}
    \caption{  {Average max-min EE versus number of iterations ($P=10$ dB, $N_{{BS}}=N_{u}=3$, $P_c=3$ W, $M=100$, and $K=4$).}}
	\label{Fig-11}  
\end{figure}
Fig. \ref{Fig-11} illustrates the convergence behavior of the algorithms considered by plotting the max-min EE achieved versus the number of iterations. The figure characterizes the practical optimality-complexity trade-off inherent in iterative optimization methods. As observed, the proposed RIS-RSMA scheme rapidly improves the objective value and exceeds the final converged EE of all benchmark strategies within as few as six iterations. This result indicates that a performance superior to the benchmark schemes can be achieved with a small number of iterations, which is particularly relevant for latency-sensitive implementations. Although RIS-assisted optimization involves higher per-iteration complexity than No-RIS schemes, the fast convergence significantly mitigates the overall computational burden. Moreover, when channel variations are slow, the optimized beamforming and RIS configurations can be reused across multiple time slots, rendering the associated computational overhead negligible from a system-level latency perspective.

\section{Conclusions}\label{sec-v}
The EE-latency trade-off of the MU-MIMO RIS-aided URLLC RSMA DL was investigated. RSMA substantially reduces the maximum delay while simultaneously enhancing the EE, especially under high user load. Our results reveal that RSMA is capable of reducing the latency by up to $13\%$ of that achieved by SDMA. Moreover, RSMA and RIS exhibit a synergistic relationship, with their combined benefits increasing under stricter reliability requirements. In particular, RSMA yields higher gains in RIS-aided systems, and increasing the RIS size further amplifies these gains relative to SDMA, without altering the fundamental EE-latency trade-off. Furthermore, the latency benefits of RIS remain modest without RSMA. This highlights the complementary roles of the two technologies: RIS improves link quality, while RSMA effectively manages interference. Finally, we showed that the EE may degrade significantly when latency minimization is prioritized. Future work may extend this framework to incorporate end-to-end latency components beyond transmission delay, as well as imperfect or statistical channel state information.

\bibliographystyle{ieeetr}
\bibliography{ref2}

\begin{IEEEbiography}[{\includegraphics[width=1in,height=1.25in,clip,keepaspectratio]{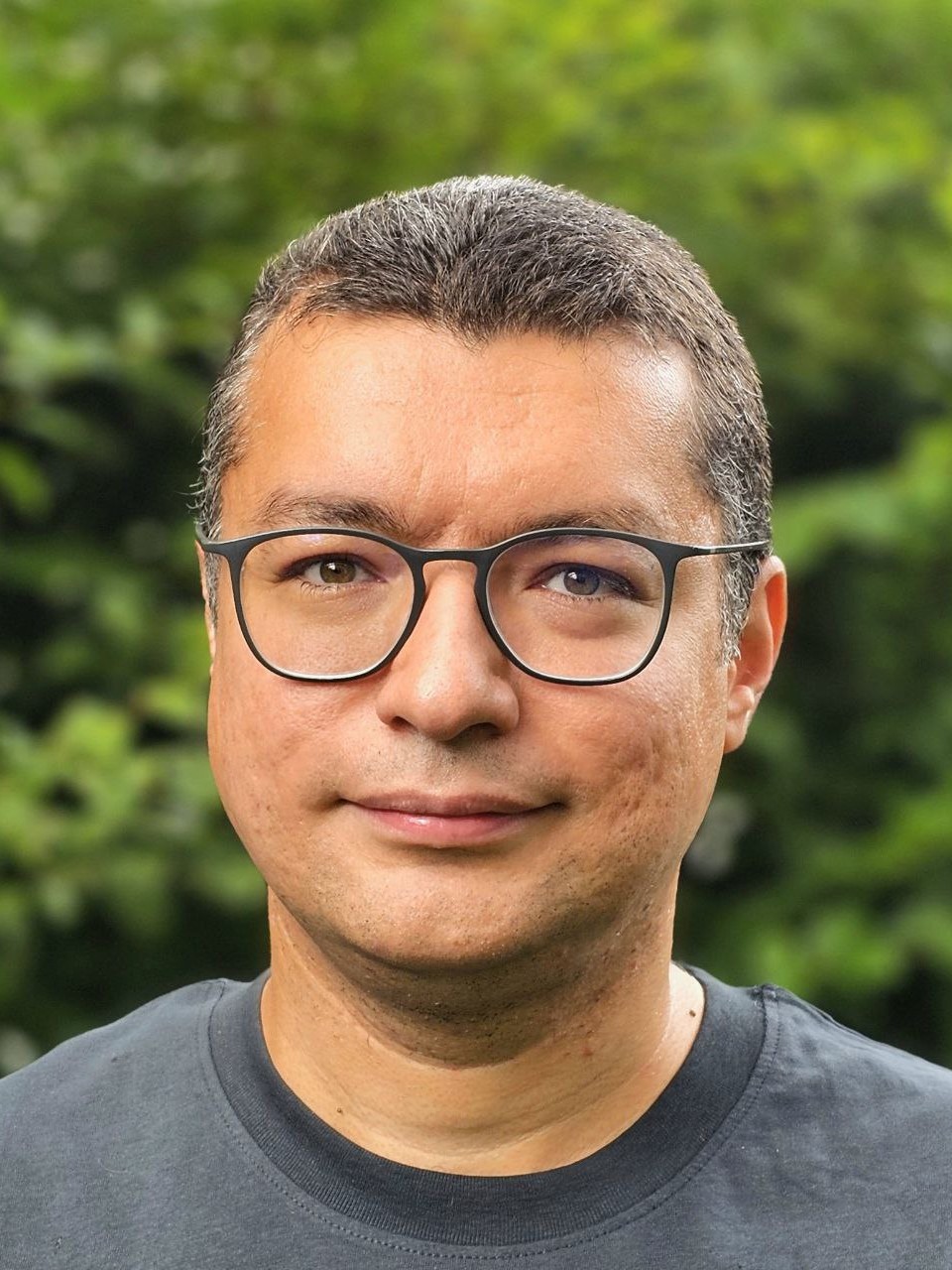}}]
{Mohammad Soleymani}  (Senior Member, IEEE) was born in Arak, Iran. He received the B.Sc. degree from Amirkabir University of Technology (Tehran Polytechnic), the M.Sc. degree from Aryamehr (Sharif) University of Technology, Tehran, Iran, and the Ph.D. degree (with distinction) in electrical engineering from the University of Paderborn, Germany. 

He is currently an Akademischer Rat a. Z. with the Signal and System Theory Group at the University of Paderborn. He was a Visiting Researcher with the University of Cantabria, Santander, Spain. His research interests include wireless communications, communication theory, signal processing, and optimization methods for next-generation communication systems, with particular emphasis on reconfigurable intelligent surfaces (RIS), multi-user MIMO, multiple access techniques, and ultra-reliable low-latency communications (URLLC).

Dr. Soleymani serves as an Editor for IEEE Communications Letters, a Handling Editor for Signal Processing (Elsevier), and an Associate Editor for the EURASIP Journal on Wireless Communications and Networking and Wireless Personal Communications. He has been an active member of the technical program committees of major conferences, including IEEE ICC, IEEE GLOBECOM, and EuCNC/6G Summit.
\end{IEEEbiography}

\begin{IEEEbiography}[{\includegraphics[width=1in,height=1.25in,clip,keepaspectratio]{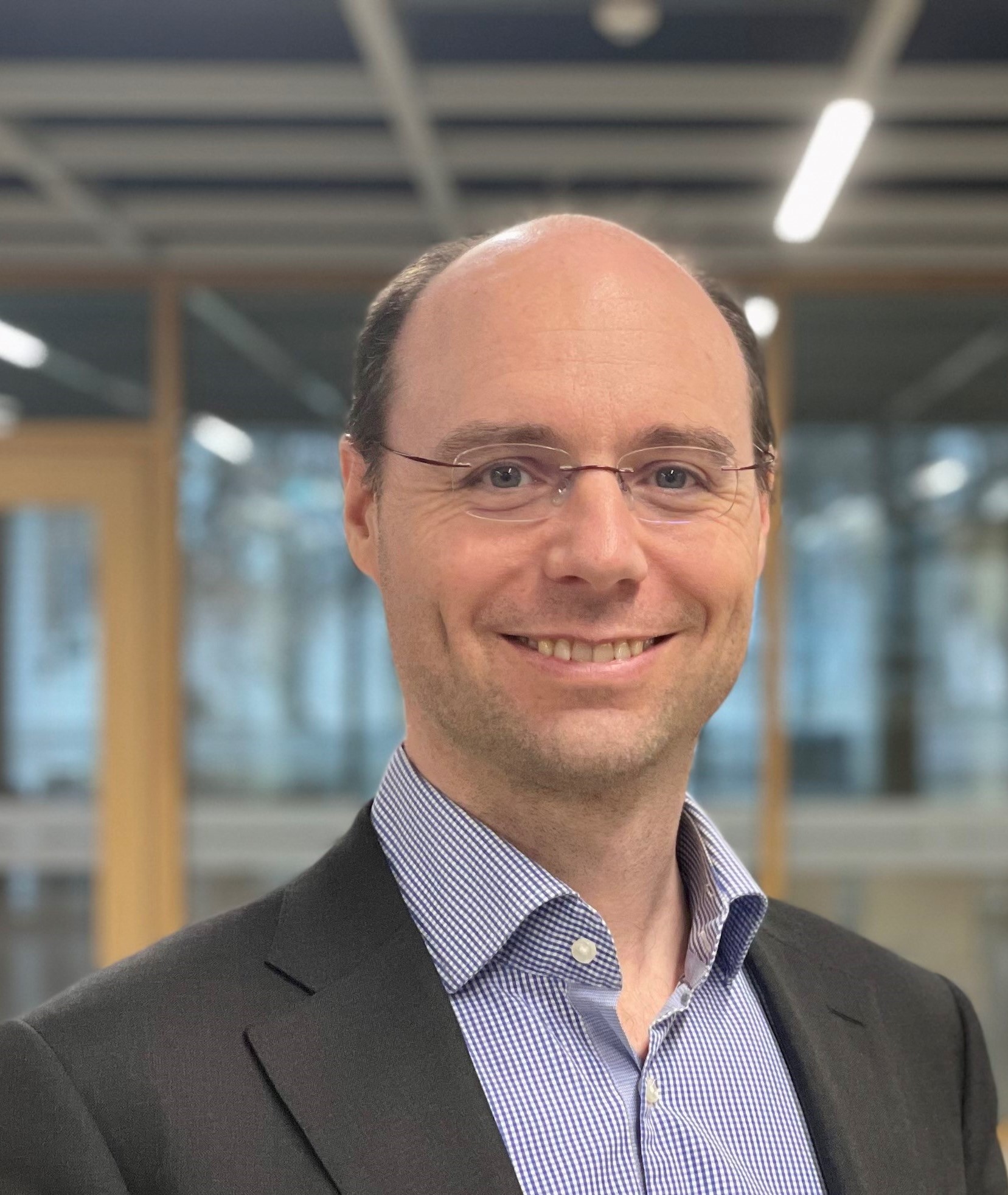}}]
{Bruno Clerckx} (Fellow, IEEE) received the M.Sc. and Ph.D. degrees in electrical engineering from Université Catholique de Louvain, Belgium, and the Doctor of Science (D.Sc.) degree from the Imperial College London, U.K. He is currently a Full Professor, the Head of the Wireless Communications and Signal Processing Laboratory, and the Deputy Head of the Communications and Signal Processing Group, Electrical and Electronic Engineering Department, Imperial College London, London, U.K. He is also the Chief Technology Officer (CTO) of the Silicon Austria Laboratories (SAL), where he is responsible for all research areas of Austria’s top research center for electronic-based systems. Prior to joining the Imperial College London in 2011, he was with Samsung Electronics, Suwon, South Korea, where he actively contributed to 4G (3GPP LTE/LTE-A and IEEE 802.16m). He has authored two books MIMO Wireless Communications and MIMO Wireless Networks, 250 peer-reviewed international research papers and 150 standards contributions, and is the inventor of 80 issued or pending patents among which several have been adopted in the specifications of 4G standards and are used by billions of devices worldwide. His research spans the general area of wireless communications and signal processing for wireless networks.

He is a fellow of the IET and an IEEE Communications Society Distinguished Lecturer. He received the prestigious Blondel Medal 2021 from
France for exceptional work contributing to the progress of Science and Electrical and Electronic Industries, the 2021 Adolphe Wetrems Prize in mathematical and physical sciences from Royal Academy of Belgium, multiple awards from Samsung, IEEE Best Student Paper Award, and the European Association for Signal Processing (EURASIP) Best Paper Award 2022. He is an IEEE Communications Society Distinguished Lecturer.

\end{IEEEbiography}

\begin{IEEEbiography}[{\includegraphics[width=1in,height=1.25in,clip,keepaspectratio]{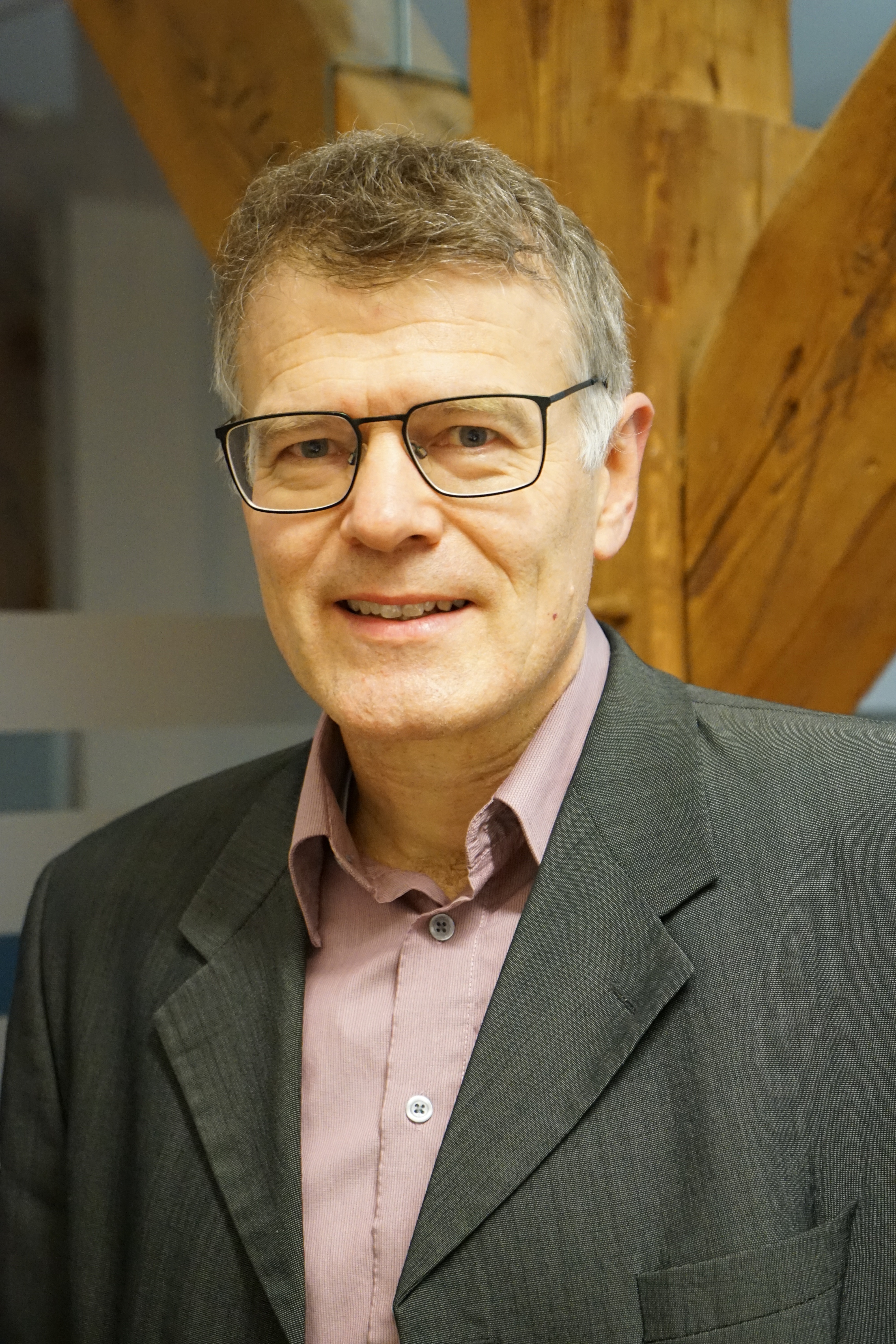}}]
{Robert Schober} (Fellow, IEEE) received the Diploma and Ph.D. degrees in electrical engineering from the Friedrich-Alexander University of Erlangen-Nuremberg (FAU), Germany, in 1997 and 2000, respectively.

From 2002 to 2011, he was a Professor and a Canada Research Chair with the University of British Columbia, Vancouver, Canada. Since January 2012, he has been an Alexander von Humboldt Professor and the Chair for Digital Communication with FAU. His research interests fall into the broad areas of communication theory, wireless and molecular communications, and statistical signal processing.

Prof. Schober received several awards for his work including the 2002 Heinz Maier Leibnitz Award of the German Science Foundation, the 2004 Innovations Award of the Vodafone Foundation for Research in Mobile Communications, the 2006 UBC Killam Research Prize, the 2007 Wilhelm Friedrich Bessel Research Award of the Alexander von Humboldt Foundation, the 2008 Charles McDowell Award for Excellence in Research from UBC, the 2011 Alexander von Humboldt Professorship, the 2012 NSERC E.W.R. Stacie Fellowship, the 2017 Wireless Communications Recognition Award by the IEEE Wireless Communications Technical Committee, the 2022 IEEE Vehicular Technology Society Stuart F. Meyer Memorial Award, and a Honorary Doctorate from Aristotle University of Thessaloniki, Greece, in 2024. Furthermore, he received numerous Best Paper Awards for his work including the 2022 ComSoc Stephen O. Rice Prize and the 2023 ComSoc Leonard G. Abraham Prize. Since 2017, he has been listed as a Highly Cited Researcher by the Web of Science. He is a Fellow of the Canadian Academy of Engineering and the Engineering Institute of Canada as well as a Member of the European Academy of Sciences and Arts, Academia Europaea, the German National Academy of Science and Engineering (acatech), and the German National Academy of Science Leopoldina. He served as an Editor-in-Chief for the IEEE Transactions on Communications, VP Publications of the IEEE Communication Society (ComSoc), and ComSoc President. He currently serves as a Senior Editor for Proceedings of the IEEE and ComSoc Past-President.

\end{IEEEbiography} 

\begin{IEEEbiography}[{\includegraphics[width=1in,height=1.25in,clip,keepaspectratio]{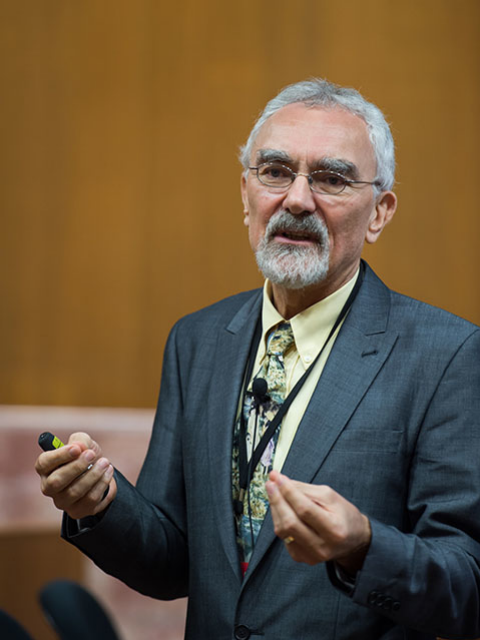}}]
{Lajos Hanzo}   (\protect\url{http://www-mobile.ecs.soton.ac.uk}, \protect\url{https://en.wikipedia.org/wiki/Lajos_Hanzo}) is a Fellow of the Royal Academy of Engineering, FIEEE, FIET, Fellow of EURASIP and a Foreign Member of the Hungarian Academy of Sciences. He coauthored 2000+ contributions at IEEE Xplore and 20 Wiley-IEEE Press monographs. He was bestowed upon the IEEE Eric Sumner Technical Field Award.

\end{IEEEbiography}  
\end{document}